\documentclass[a4paper,12pt]{article}
\usepackage{natbib}
\usepackage{amsmath} 
\usepackage[dvips]{graphicx}
\usepackage{tabularx}
\usepackage{fancyhdr}
\usepackage{setspace}
\usepackage{url}
\usepackage{enumerate}

\graphicspath{{../Dkk1_images/}}

\newcommand{\captionfonts}{\small}

\makeatletter  % Allow the use of @ in command names
\long\def\@makecaption#1#2{%
  \vskip\abovecaptionskip
  \sbox\@tempboxa{{\captionfonts #1: #2}}%
  \ifdim \wd\@tempboxa >\hsize
    {\captionfonts #1: #2\par}
  \else
    \hbox to\hsize{\hfil\box\@tempboxa\hfil}%
  \fi
  \vskip\belowcaptionskip}
\makeatother   % Cancel the effect of \makeatletter
\newcommand{\cfLA}{c_{f[LGA]}} 
 
\newcommand{\cfGA}{c_{f[GA]}} 
\newcommand{\cbGA}{c_{b[GA]}} 
\newcommand{\cfC}{c_{fC}} 
\newcommand{\cbC}{c_{bC}} 
\newcommand{\ctc}{c_{tsc}}
\newcommand{\ctl}{c_{tsl}} 
 
\newcommand{\bcatenin}{$\beta$-catenin}

\begin{document}

%Title must not exceed 100 characters
\title{Dickkopf1 - a new player in modelling the Wnt pathway} 
\author{Lykke Pedersen$^*$, Sandeep Krishna and Mogens H. Jensen \\ Niels Bohr Institute, Blegdamsvej 17, 2100 Copenhagen \O, Denmark }
%\affiliation{Niels Bohr Institute, Blegdamsvej 17, DK-2100, Copenhagen, Denmark}

\date{}

\maketitle
\begin{flushleft}
*: Corresponding author, lykkep@nbi.k, Phone:   +45 353 25260, Fax:  +45 353 25425  \\
Character count: 21955 \\
%running title at most 50 characters including spaces
Running title: Dkk1 - a new player in Modelling the Wnt pathway \\ %48 characters 
Subject Category: Development

\end{flushleft}
\newpage

%%%%%%%%%%%%%%%%%%%%%%%%%%%%%%
\section*{Abstract}
%%%%%%%%%%%%%%%%%%%%%%%%%%%%%%
%This should be a single paragraph not exceeding 175 words
%Currently: 145
The Wnt signalling pathway transducing the stabilization of \bcatenin{} is essential for metazoan embryo development and is misregulated in many diseases such as cancers. In recent years models have been proposed for the Wnt signalling pathway during the segmentation process in developing embryos. Many of these include negative feedback loops build around Axin2 regulation. In this article we propose a new negative feedback model for the Wnt pathway with Dickkopf1 (Dkk1) at its core. Dkk1 is a negative regulator of Wnt signalling. In chicken and mouse embryos there is a gradient of Wnt in the presomitic mesoderm (PSM) decreasing from the posterior to the anterior end. The formation of somites and the oscillations of Wnt target genes are controlled by this gradient. Here we incorporate a Wnt gradient and show that synchronization of neighbouring cells in the PSM is important 
in accordance with experimental observations. \\

% Up to five keywords
Keywords: Clock and wavefront model / Dkk1 / Modelling / Segmentation / Wnt pathway 

%%%%%%%%%%%%%%%%%%%%%%%%%%%%%%
\section*{Introduction}
%%%%%%%%%%%%%%%%%%%%%%%%%%%%%%

A segmented body plan is a fundamental characteristic feature of vertebrates. The process of segmentation is carried out by a combination of changes in gene expression and relative anterior-posterior cell position in the presomitic mesoderm (PSM) \citep{WolpertBOOK}. In the anterior end of the 
embryo the somites are segmented at equally separated time points with species dependent periods. In mice the period is around $120 \, \rm{min}$ and in frogs it is around $90 \, \rm{min}$. 

In \citeyearpar{Cooke1976} Cooke and Zeeman proposed the clock and wavefront model to describe the segmentation process. The idea is that locally coupled oscillators are controlled by a morphogen gradient in the PSM. The oscillators are the clocks providing temporal information, e.g., cycle state, and the morphogen gradient is the wavefront providing spatial information about axial position. Until now three major pathways controlling the segmentation process have been found: the Notch, Wnt and FGF pathways. They all have target genes, which oscillates and, interestingly, Wnt target genes oscillate out of phase with Notch and FGF target genes \citep{Dequeant2006}. These three pathways could be the clocks. There is a decreasing gradient of wnt3a and fibroblast growth factor 8 (fgf8) starting from the tail bud through the PSM \citep{Aulehla2003,Dubrulle2004a}. Fgf8 seems to be regulated by Wnt \citep{Aulehla2004}, thus Wnt could form the main wavefront.
The actual setting of the somites happens at the determination front, where the gradient drops below a certain threshold. Cells past the determination front become permissive to form somites depending on their phase of oscillation \citep{Aulehla2004, Pourquie2004}. 

In 2003 it was discovered by Aulehla et al. that Axin2 oscillates during the segmentation process in developing mouse embryos. Since their discovery several models for the Wnt oscillator have been proposed \citep{Goldbeter2008,Lee2003a,Jensen2010} with Axin2 as the main variable. We propose a new model for the Wnt pathway with Dickkopf1 (Dkk1) as the core of a negative feedback loop. The reason for changing from Axin2 to Dkk1 is that Axin2 deleted mice (Axin2$^{-}/^{-}$) have no dramatic phenotype \citep{Yu2005}. Dkk1 has an oscillatory behavior during the segmentation process in mouse embryos \citep{Dequeant2006} and lack of Dkk1 in mouse embryos results in anterior head truncations and neonatal death \citep{MacDonald2004}. In mice, for which we want to model the segmentation process, several Dickkorpf genes are known. Dkk1 is the most widely studied of these.

%%%%%%%%%%%%%%%%%%%%%%%%%%%%%%%
\section*{Results and Discussion} 
%%%%%%%%%%%%%%%%%%%%%%%%%%%%%%

%%%%%%%%%%%%%%%%%%%%%%%%%%%%%%
\subsection*{Modeling the Wnt/\bcatenin{} pathway}

During Wnt signaling \bcatenin{} interacts with the TCF/LEF-1 DNA-binding proteins to promote transcription of Wnt target genes \citep{Logan2004}. As for Axin2 the transcription factor for Dkk1 is $\beta$-catenin \citep{Niida2004,GonzalezSancho2005}. After transcription and translation Dkk1 goes through the cellular membrane where it can bind to the extracellular domains of the low-density lipoprotein receptor-related protein 5 and 6 (LRP5/6). When bound to LRP5/6, Dkk1 acts as an inhibitor of Wnt signaling by blocking the association between Wnt, Frizzled (Fz) and LRP5/6 \citep{Semenov2001}. Wnt acts as an inducer for the formation of this complex and Dkk1 is a competitor to this induction \citep{MacDonald2009,Tamai2000}. 

It has been proposed that the Wnt signal is transduced through the cell membrane by the binding of Dishevelled (Dsh) to the intracellular domain of the Fz receptor\citep{Wallingford2005}. Axin and Dsh can bind together via their DIX domains \citep{Wallingford2005} and they co-localize at the membrane \citep{Fagotto1999} during Wnt signaling. Therefore Dsh bound to Fz may recruit Axin bound to GSK3$\beta$ to the LRP5/6 receptor \citep{Huang2008}, where a phosphorylation of LRP5/6 is initiated. The LRP5/6 receptor has a binding site for Axin and upon Wnt signalling GSK3$\beta$ (bound to Axin) phosphorylates LRP5/6, which requires Axin \citep{Huang2008}. The phosphorylated LRP5/6 receptor may be able to recruit and more efficiently bind the Axin-GSK3$\beta$ complex to the membrane and the phosphorylation process is thereby amplified \citep{MacDonald2009}.  

At the cell membrane Axin is phosphorylated by GSK3$\beta$ and then degraded \citep{Mao2001b,Yamamoto1999}. The degradation of Axin leads to a decrease in the formation of the destruction complex comprised of $\beta$-catenin, the kinases glycogen synthase kinase 3 (GSK3$\beta$), casein kinase (CKI$\alpha$), the scaffolding proteins Axin and adenomatous polyposis coli (APC). In the destruction complex \bcatenin{} gets phosphorylated and subsequently degraded. 

Interestingly enough GSK3$\beta$ plays a dual role in controlling the Wnt signal. When the Wnt signal is off then GSK3$\beta$ phosphorylates \bcatenin{} in the destruction complex and when the signal is on then it phosphorylates Axin at the LRP5/6 receptor. Whether it is the same or distinct Axin-GSK3$\beta$ complexes that carry out the phosphorylation of \bcatenin{} and Axin is unknown \citep{Huang2008}.   \\

The dynamics of CK1$\alpha$ and APC are included in the parameters governing the destruction complex, the TCF/LEF-1 dynamics are contained in the transcription of Dkk1, and the dynamics of Dsh are included in the formation of the complex consisting of Axin, GSK3$\beta$ and LRP5/6 at the cell membrane. Figure \ref{fig:modeldiagram} shows a simplified diagram of the proposed model 
and the associated equations. 
$C$, $[GA]$, $G$, $B$, $L$, $D_m$, $D$, $[LD]$, $A$ and $[LGA]$ are the concentrations of the destruction complex, GSK3$\beta$-Axin complex, GSK3$\beta$, \bcatenin{}, LRP5/6, Dkk1 mRNA, Dkk1 protein, Dkk1-LRP5/6 complex, Axin and LRP5/6-Axin-GSK3$\beta$ complex. The formation and breaking of a complex $X$ are denoted by $c_{fX}$ and $c_{bX}$, respectively. The transcription and translation rate of Dkk1 are given by the parameters $\ctc$ and $\ctl$, respectively. The Hill coefficient on $B$ regarding the transcription of $D_m$ is associated to the amount of cooperativity between \bcatenin{} and the TCF/LEF-1 complex. For example, no cooperativity would result in a Hill coefficient of one. 

The concentration of GSK3$\beta$ has been shown to be extremely stable \citep{Lee2003a} and consequently its total concentration, $GSK3\beta_{tot}$, is assumed to be constant during the involved time scales. The same assumption goes for the total concentration of LRP5/6, since the half-life of LRP6 is around 4.7 hours \citep{Semenov2008}. Therefore we have not included any source or sink for the concentrations of $G$ and $L$. Only constitutive sources of \bcatenin{} and Axin, respectively, are included in the model, because free (unphosphorylated) \bcatenin{} and Axin is stable \citep{Lee2003a}. 

%%%%%%%%%%%%%%%%%%%%%%%%%%%%%%
\subsection*{Determination of the parameter values} 
In \citet{Lee2003a} a model for the Wnt pathway in \textit{Xenopus} is proposed and from this article estimates for the parameters $K_C=\frac{\cbC+\alpha}{\cfC}$, $K_{[GA]}=\frac{\cbGA}{\cfGA}$, $\alpha$, GSK3$\beta_{tot}$, $S_A$ and $S_B$ are taken. Dkk1 binds to LRP5/6 with a high affinity, since the dissociation constant has been measured to be around $K_{[LD]}=0.4 - 0.5  \,\rm{nM}$ \citep{Semenov2001,Bafico2001}. 

The other parameters are estimated to produce oscillations with a period of around $120 \, \rm{min}$ and a very low concentration of total Axin as found by \citet{Lee2003a}. The low concentration of total Axin is thought to act as a buffer to changes in the concentration of the other constituents, which may also take part in other signalling pathways \citep{Logan2004}. 

In the activation of transcription of Dkk1 by \bcatenin{} we assume cooperativity between \bcatenin{} and TCF/LEF. The Hill coefficient is set to three. A Hill coefficient of one and two returns no sustained oscillations, when the model is simulated. In other models of the Wnt pathway Hill coefficients of values 2 and 5 have been applied \citep{Lee2003a,Wawra2007}.

%%%%%%%%%%%%%%%%%%%%%%%%%%%%%%
\subsection*{Oscillations of the mRNA, protein and complex levels} 
The parameter values produce oscillations of the involved constituents with a period of around $120 \, \rm{min}$ as seen in Fig.~\ref{fig:modeldiagram}. In addition, the concentration of Axin is very low compared to the concentrations of the other variables, as discussed above. The different steps in the proposed Wnt model comply with the phase shifts in the oscillations of the different components. An increase in the $[LGA]$ concentration leads to a decrease in the Axin concentration. This decrease causes a reduction in the concentration of the destruction complex and consequently an increase in the \bcatenin{} concentration. This increase will after a while cause an increase in the Dkk1 concentration. The concentrations of GSK3$\beta$ and $[GA]$ are mirrors of each other, since a high concentration of
will cause $[GA]$ to leave less free GSK3$\beta$ behind. 

\citet{Aulehla2008} found no significant oscillations in the level of \bcatenin{}. For our choice of parameters \bcatenin{} show oscillatory behavior with an amplitude of approximately $5 \, \rm{nM}$, which is not significantly low. A different set of parameters could possibly give a smaller amplitude of \bcatenin{} but the general results, presented later, is not significantly altered by this. 

%%%%%%%%%%%%%%%%%%%%%%%%%%%%%%
\subsection*{A spatial Wnt gradient induced by time variation}
The described model could be the clock of the clock and wavefront model as mentioned above. The wavefront we introduce as a time dependent Wnt gradient through a time dependence of the parameter $\cfLA$. The reason for making 
$\cfLA$ time dependent is clear if we introduce the variable $[LW]$ describing the binding of Wnt ($W$) to the LRP5/6 receptor together with Frizzled. The temporal change of $[LW]$ is described as 
\begin{equation}
\frac{d [LW]}{dt} = c_{f[LW]}LW(t)-c_{b[LW]}[LW] \enspace,
\end{equation}
and if we assume steady state for the binding of Wnt to LRP5/6 then 
\begin{equation}
[LW]=\frac{c_{fLW} LW(t)}{c_{b[LW]}}=\frac{LW(t)}{K_{[LW]}} \enspace. 
\end{equation}
Inserting this $L$ estimate into the concentration of $[LGA]$, it can be seen that the formation of $[LGA]$ will be time dependent, with 
\begin{equation}
\cfLA(t)=\frac{\cfLA W(t)}{K_{[LW]}} \enspace. 
\end{equation}
To include the decreasing gradient of Wnt observed in mouse embryos we use a decreasing value of $\cfLA$. If we assume that Wnt primarily decreases through a diffusive process, then the profile of $\cfLA$ will be Gaussian. In Fig.~\ref{fig:cfLGA} the amplitude and period of Dkk1 concentration is plotted as $\cfLA$ changes. The green line refers to the reference value of $\cfLA$; reducing it results in smaller amplitudes. The period seems to be almost unaltered, which complies with the findings of constant periods of the oscillations in the PSM.

The length of the PSM is approximately constant during the formation of the first 15-20 somites in mice embryos \citep{Tam1981}. The same is almost true for the size of the somites. From \citet{Tam1981} one can find that a cell budded off in the tail bud will be segmented within the order of $\sim 1100-1300 \, \rm{mins}$.

 It has been measured that FGF (regulated by Wnt) exhibits a gradient in the PSM with a fold change of 2-5 \citep{Dubrulle2004a}. The fold change used for $\cfLA$ is $2$. Assuming that Wnt diffuses through the PSM and setting the final value of $\cfLA$ equal to $2.5 \frac{1}{\rm{nMmin}}$ enable us to calculate a Gaussian profile of $\cfLA$ representing the Wnt gradient. In Supplementary Fig. 1, Gaussian profiles of $\cfLA$ are plotted with different initial values. A decreasing value of $\cfLA$ in the PSM will give rise to smaller amplitudes and slightly longer periods.

%%%%%%%%%%%%%%%%%%%%%%%%%%%%%%
\subsection*{Synchronization of neighbouring cells}

The elongation of the embryo is modelled by simulating cells budding off from the tail bud at regular time intervals, $R$. Elongation is considered in only one dimension. In Fig.~\ref{fig:3D} time series for cells with $R=10 \, \rm{min}$ are shown. The final level of Dkk1 at the determination front is oscillating (see Fig.~\ref{fig:3D}A) with a period of $120 \, \rm{min}$. However, these oscillations are only possible if the onset of oscillations in the cells are synchronized, i.e. if the state for cell $i$ at time $t$ is denoted $S_{i}(t)$, then $S_{i+1}(0)=S_i(R)$. This means that the initial state of a cell budded off from the tail bud equals that of its anterior neighbour. If all the cells have the same initial state then the levels of Dkk1 at the determination front will also be equal, see Fig.~\ref{fig:3D}B. 

One could imagine that the synchronization of the clocks in each cell is not perfect. If the initial state of cell $i+1$ is chosen randomly as $S_{i+1}(0)=S_i(R+\mathcal{U}(-\frac{R}{2},\frac{R}{2}))$, then the period is almost unaltered (see Supplementary Fig. 2A). Here $\mathcal{U}(-x,x)$ denotes random numbers uniformly distributed between $-x$ and $x$. If the initial state is randomly chosen within the range of Dkk1 levels, then the oscillations of Dkk1 at the determination front are disrupted and no periodicity is visual (see Supplementary Fig. 2B). Thus, the system seems to be robust to small changes in the synchronization.

Diffusion of Wnt is thought to be the main reason for the Wnt gradient in the PSM. Thus a Gaussian profile of $\cfLA$ is used. The profile of $\cfLA$ would still be Gaussian if we also included a half-life of Wnt. If the Wnt gradient was only controlled by the half-life of Wnt then the levels of Dkk1 still oscillate at the determination front with a period of 120 min, when the model is simulated as above with $R=10 \, \rm{min}$. %Since \citet{Aulehla2008} found a decreasing gradient of \bcatenin{} in the PSM, $S_B$ was given a Gaussian profile in the PSM with a fold change of 2 from the tail bud to the anterior part of the PSM. Still the oscillations of Dkk1 at the determination front have a period of 120 min, when the model is simulated as above with $R=10 \, \rm{min}$.

%%%%%%%%%%%%%%%%%%%%%%%%%%%%%%
\subsection*{The period of somite formation increases}
It is known that the period of somite formation increases as the embryo elongates \citep{Tam1981}. This led \citet{Aulehla2004} to propose that the level of Wnt production in the tail bud increases with time. To test their hypothesis we let the initial value of $\cfLA$ increase with time keeping the level at the determination front fixed at $2.5 \frac{1}{\rm{nMmin}}$ as above. Simulating an increase in the Wnt level of the tail bud results in longer periods of the Dkk1 levels at the determination front, see Supplementary Fig. 3. Thus, we can confirm the hypothesis of \citet{Aulehla2004}. Note that a steeper increase of the Wnt level in the tail bud results in a greater extension of the periods (compare Supplementary Figs. 3A and 3B). 

%%%%%%%%%%%%%%%%%%%%%%%%%%%%%%
\section*{Outlook}
%%%%%%%%%%%%%%%%%%%%%%%%%%%%%%
We propose a Wnt model with Dkk1 as the core for the negative feedback loop, which exhibits sustained oscillations of Dkk1. By simulation of the elongating embryo we were able to test the importance of synchronization between neighbouring cells. In addition, we could also show that a small change in the synchronization did not significantly disrupt smooth oscillations of Dkk1 levels at the determination front. The clock and wavefront model was investigated through a decreasing Wnt gradient in the PSM and we could confirm the hypothesis that the Wnt level in the tail bud increases over time resulting in longer periods of the oscillations. 
Further aspects of the growing embryo could be incorporated into the model, like a varying PSM size and a rate of cells  budding off from the tail bud.

%%%%%%%%%%%%%%%%%%%%%%%%%%%%%%
\section*{Acknowledgments}
%%%%%%%%%%%%%%%%%%%%%%%%s%%%%%%
This research is supported by the Danish Research Foundation.

\bibliographystyle{npg}
\bibliography{/Users/lykkep/Documents/PhD/Thesis/PhDThesis}

\begin{thebibliography}{}

\bibitem[\protect\citename{Goldbeter \& Pourqui{\'e}, }2008]{Goldbeter2008}
Goldbeter, A, \& Pourqui{\'e}, O. 2008.
\newblock Modeling the segmentation clock as a network of coupled oscillations
  in the Notch, Wnt and FGF signaling pathways.
\newblock {\em J Theor Biol}, {\bf 252}(3), 574--585.

\end{thebibliography}

\newpage

%%%%%%%%%%%%%%%%%%%%%%%%%%%%%%
\section*{Figure legends}
%%%%%%%%%%%%%%%%%%%%%%%%%%%%%%
%1
\begin{figure}[!h]
\caption{\textbf{Top,left}: A diagram of the Wnt model with a feed-back loop over Dkk1. Included are only members of the Wnt pathway that are important for the understanding of the negative feedback loop. When the Wnt signal is on, then Axin gets degraded at the LRP5/6 complex and \bcatenin{} can act as a transcription factor of the Wnt inhibitor Dkk1. Vice-versa, when the Wnt signal is off, due to inhibition by Dkk1, then \bcatenin{} gets degraded. \textbf{Top,right} Simulated time series for a selection of variables from the model listed in the bottom panel. The total level of Axin (magenta) is low, which complies with the findings of \citet{Lee2003a}. \textbf{Bottom}: The terms describing the time dependence of the variables. } \label{fig:modeldiagram} 
\end{figure}

%2
\begin{figure}[!h]
\caption{\textbf{Top}: Time series for the Dkk1 concentration. Space is introduced by letting a cell bud off from the tail bud every 10th min. Thus the cells move relatively in the PSM. At the determination front the oscillations arrest. \textbf{Bottom}: The level of Dkk1 at the determination front. (A) $R = 10 \, \rm{min}$. (B) $R = 10 \, \rm{min}$. (C) $R = 10 \, \rm{min}$ and the cells have the same initial level of Dkk1
TB: tail bud. S$_{i}$: Somite $i$, where S$_{0}$ is the newly formed somite. }\label{fig:3D} 
\end{figure}

%3
\begin{figure}[!h]
\caption{The parameter $\cfLA$ is changed, and the amplitude (solid) and the period (dashed) are calculated. The green vertical line denotes the value of $\cfLA$ in the reference state.} \label{fig:cfLGA}
\end{figure}

\newpage

%%%%%%%%%%%%%%%%%%%%%%%%%%%%%%
%\section*{Figures}
%%%%%%%%%%%%%%%%%%%%%%%%%%%%%%

\subsection*{Figure 1}
\begin{center}
\includegraphics[width=0.9\textwidth]{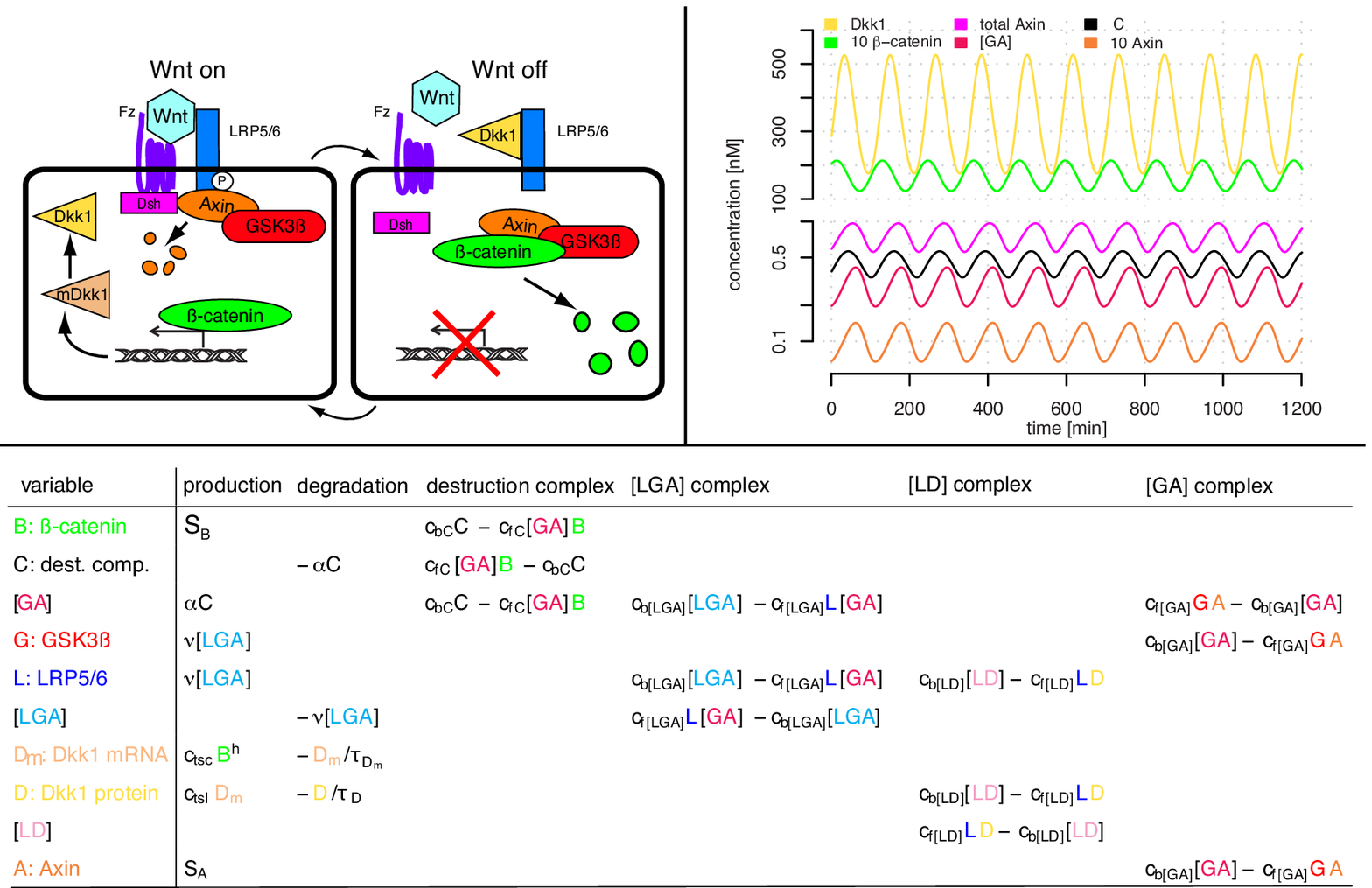}
\end{center}

\subsection*{Figure 2}
\begin{center}
\includegraphics[width=\textwidth]{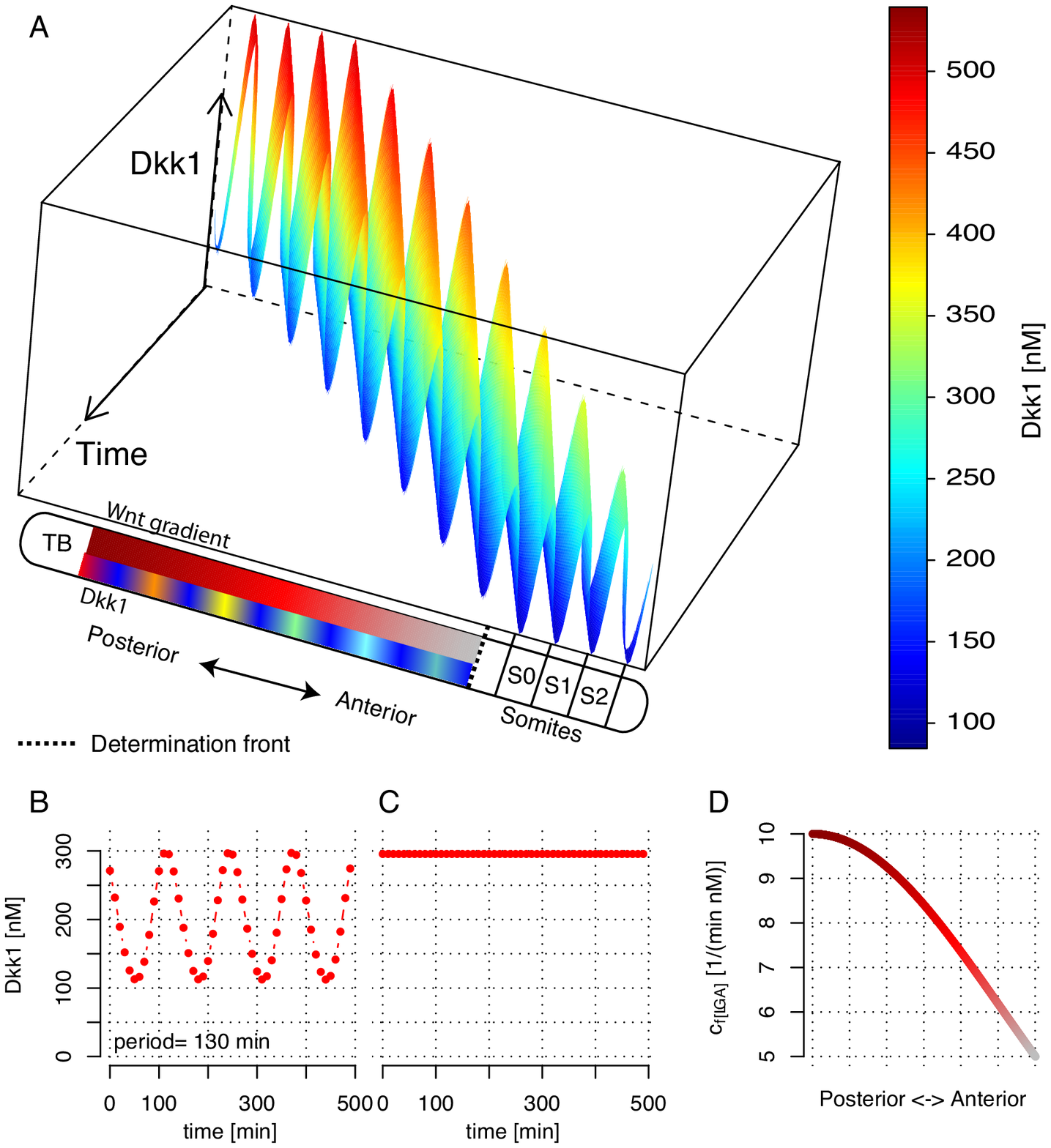}
\end{center}

\subsection*{Figure 3}
\begin{center}
\includegraphics[angle=-90,width=\textwidth]{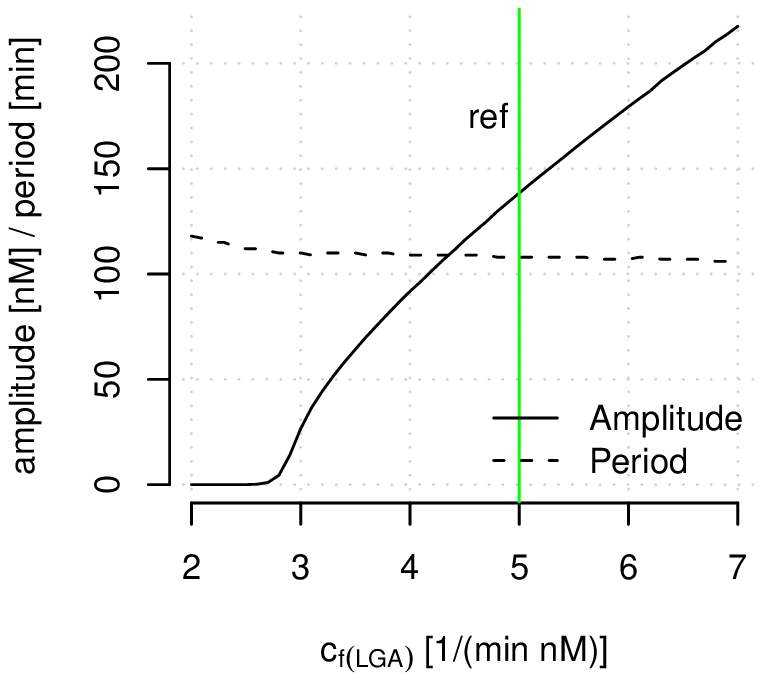}
\end{center}

\newpage

%%%%%%%%%%%%%%%%%%%%%%%%%%%%%%
\section*{Tables}
%%%%%%%%%%%%%%%%%%%%%%%%%%%%%%
\begin{table}[!h] 
\caption{Parameters in our model of the Wnt system and their default values}\label{tb:par}
\setlength\extrarowheight{3pt}  %Increases the height of each row
\addtolength{\tabcolsep}{5pt}  %Increases the distance between columns
\begin{tabularx}{\linewidth}{lll} 
\hline
Parameter & Process & Default Value \\
\hline 
$K_{C}$		& Dissociation constant $C$		  		&  $8 \, \rm{nM}$\\ 
$c_{bC}$		& Breaking of $C$						&  $7 \, \rm{min}^{-1}$\\ 
$\alpha$    	& Degradation of $\beta$-catenin 			&  $2.2 \, \rm{min}^{-1}$\\ 
$K_{[GA]}$ 	& Dissociation constant $[GA]$			  	&  $1.5 \, \rm{nM}$\\ 
$c_{b[GA]}$ 	& Breaking of $[GA]$ 					&  $4 \, \rm{min}^{-1}$\\
$K_{[LGA]}$	& Dissociation constant $[LGA]$			& $1 \, \rm{nM}$\\
$c_{b[LGA]}$	& Breaking of $[LGA]$					&  $10 \, \rm{min}^{-1}$\\
$\nu$		& Degradation of Axin					& $3.8 \, \rm{min}^{-1}$ \\
$K_{[LD]}$	& Dissociation constant $[LD]$				& $0.5 \,\rm{nM}$ \\
$c_{b[LD]}$	& Breaking of $[LD]$ 					&  $0.02 \, \rm{min}^{-1}$\\ 
$S_B$ 		& Constant source of $\beta$-catenin  		&  $1 \, \rm{nM/min}$\\ 
$S_A$ 		& Constant source of Axin 				&  $0.02 \, \rm{nM/min}$\\ 
$c_{tsl}$  		& Transcription of \textit{dkk1}				&  $0.02 \, \rm{min}^{-1}$\\ 
$c_{tsc}$ 	 	& Translation of Dkk1 mRNA				&  $0.025 \, \rm{(Nm^2 \, min)}^{-1}$\\ 
$\tau_{Dm}$ 	& Average lifetime of dkk1 mRNA 			& $8 \, \rm{min}$\\
$\tau_{D}$ 	& Average lifetime of Dkk1 				& $16 \, \rm{min}$\\
\hline 
$GSK3\beta_{tot}$ & Total G level 						&  $45 \,  \rm{nM}$ \\
$L_{tot}$ 		& Total L level							& $15 \, \rm{nM}$ \\
\hline
\end{tabularx} \label{Wnt rates}

\end{table}

\end{document}